\documentclass[aps,prl,superscriptaddress,showpacs,preprint]{revtex4}
\usepackage{bm,graphicx,graphics,amsmath,amssymb,bm,epsfig,color}
\usepackage{longtable}
\usepackage{epstopdf}

\begin{document}

\title{Time-Resolved Studies of the Spin-Transfer Reversal Mechanism in Perpendicularly Magnetized Magnetic Tunnel Junctions}

\author{Christian Hahn}
\affiliation{Department of Physics, New York University, New York, New York 10003, USA}
\author{Georg Wolf}
\affiliation{Department of Physics, New York University, New York, New York 10003, USA}
\affiliation{Spin Transfer Technologies, Inc., Fremont, California 94538, USA}
\author{Bartek Kardasz}
\affiliation{Spin Transfer Technologies, Inc., Fremont, California 94538, USA}
\author{Steve Watts}
\affiliation{Spin Transfer Technologies, Inc., Fremont, California 94538, USA}
\author{Mustafa Pinarbasi}
\affiliation{Spin Transfer Technologies, Inc., Fremont, California 94538, USA}
\author{Andrew D. Kent}
\affiliation{Department of Physics, New York University, New York, New York 10003, USA}
\date{\today}

\begin{abstract}

Pulsed spin-torque switching has been studied using single-shot time-resolved electrical measurements in perpendicularly magnetized magnetic tunnel junctions as a function of pulse amplitude and junction size in 50 to 100\;nm diameter circular junctions. The mean switching time depends inversely on pulse amplitude for all junctions studied. However, the switching dynamics is found to be strongly dependent on junction size and pulse amplitude. In 50\,nm diameter junctions the switching onset is stochastic but the switching once started, is fast; after being initiated it takes less than 2\,ns to switch. In larger diameter junctions the time needed for complete switching is strongly dependent on the pulse amplitude, reaching times less than 2\,ns at large pulse amplitudes. Anomalies in the switching rate versus pulse amplitude are shown to be associated with the long lived ($> 2$ ns) intermediate junction resistance states. 

\end{abstract}

\maketitle
Perpendicularly magnetized magnetic tunnel junctions (pMTJs) are interesting physical systems and promising for high-density, non-volatile data storage, notably magnetic random access memory (MRAM) \cite{Ikeda2010,Worledge2012,Kent2015,Nowak2016}, particularly when they are sub-100 nm in lateral scale. They are model systems to explore spin transfer (ST) induced magnetization reversal because of their uniaxial magnetic anisotropy, which favors magnetization oriented up or down along an axis perpendicular to the plane of a thin free magnetic layer. In fact, this is one of the few physical situations for which analytic models exists for the thermal stability \cite{Adam2012,Chaves2015} and ST switching time \cite{Sun2000}, the latter in a macrospin model which assumes that the entire magnetization of a layer rotates uniformly during reversal. (For a review of this model see \cite{Liu2014}.) Micromagnetic simulations have also been used to study spin-transfer switching in pMTJs \cite{Munira2015}. The basic physics of switching is associated with the interplay of ST torque due to the flow of spin-polarized electrons, magnetization damping and thermal fluctuations. Thermal fluctuations of the magnetization can lead to stochastic reversal that is assisted by ST torques. However, when the ST torque exceeds the damping torque---equivalent to a current or voltage exceeding a threshold---one enters a dynamic regime in which ST can drive the reversal. In this limit the switching can occur on nanosecond time scales and the switching time is predicted to scale inversely with the pulse amplitude
\cite{Sun2000}.

Early experiments that investigated spin-valve nanopillars with perpendicular magnetic anisotropy found that the switching time was indeed inversely proportional to the pulse amplitude \cite{Bedau2010a,Bedau2010}. However, these experiments only determined the switching probability as a function of pulse amplitude and duration, so they could not directly probe the magnetization reversal mechanism. Scanning transmission x-ray microscopy experiments on similar samples were able to both time and spatially resolve the magnetization dynamics but used a pump-probe method that averaged over many events \cite{Bernstein2011}. The advent of pMTJ with large magnetoresistance and low switching voltages \cite{Ikeda2010,Worledge2012} enables single-shot time resolved electrical measurements of spin-transfer switching.  Such techniques have been used in the past to study switching of in-plane magnetized MTJs \cite{Devolder2008,Tomita2008,Cui2010,Heindl2014,Wolf2014}. Very recently time resolved studies of pMTJs were performed in a thermally assisted ST regime by Devolder {\it et al.} \cite{Devolder2016_a,Devolder2016_w} and showed time-responses that indicate a domain wall mediated reversal of the magnetization. 

In order to elucidate the reversal mechanisms in the ST dynamic regime we performed pulsed single-shot time resolved measurements of ST-switching in circular pMTJs of varying sizes. This allows precise extraction of the time when switching occurs with respect to the turn-on of the pulse, as well as the transition time needed for the switching process. Our pulsed switching scheme involves recording the junction's resistance response after turning the voltage pulse on in an abrupt ($\mathrm{<}$100\,ps) step. With time-resolved measurements of the sample resistance during application of the pulse we acquire statistics on magnetization reversal events as a function of pulse amplitude, in the dynamic regime, in which the pulse amplitude is larger than the threshold for switching. These studies are complementary to the studies of ST thermally assisted switching induced by a voltage ramp \cite{Devolder2016_a,Devolder2016_w}.

Our samples contain a perpendicularly magnetized 1.1~nm thick CoFeB free layer that forms an MgO tunnel barrier with a synthetic antiferromagnet (SAF) reference layer grown by standard methods \cite{Sato2012,Worledge2012,Wang2015}. Circular junctions with diameters ranging from 50 to 100\,nm were patterned from a single wafer (i.e. all junctions are formed from the same multilayer, only their lateral size varies).
\begin{figure}
\includegraphics[width=0.5\textwidth]{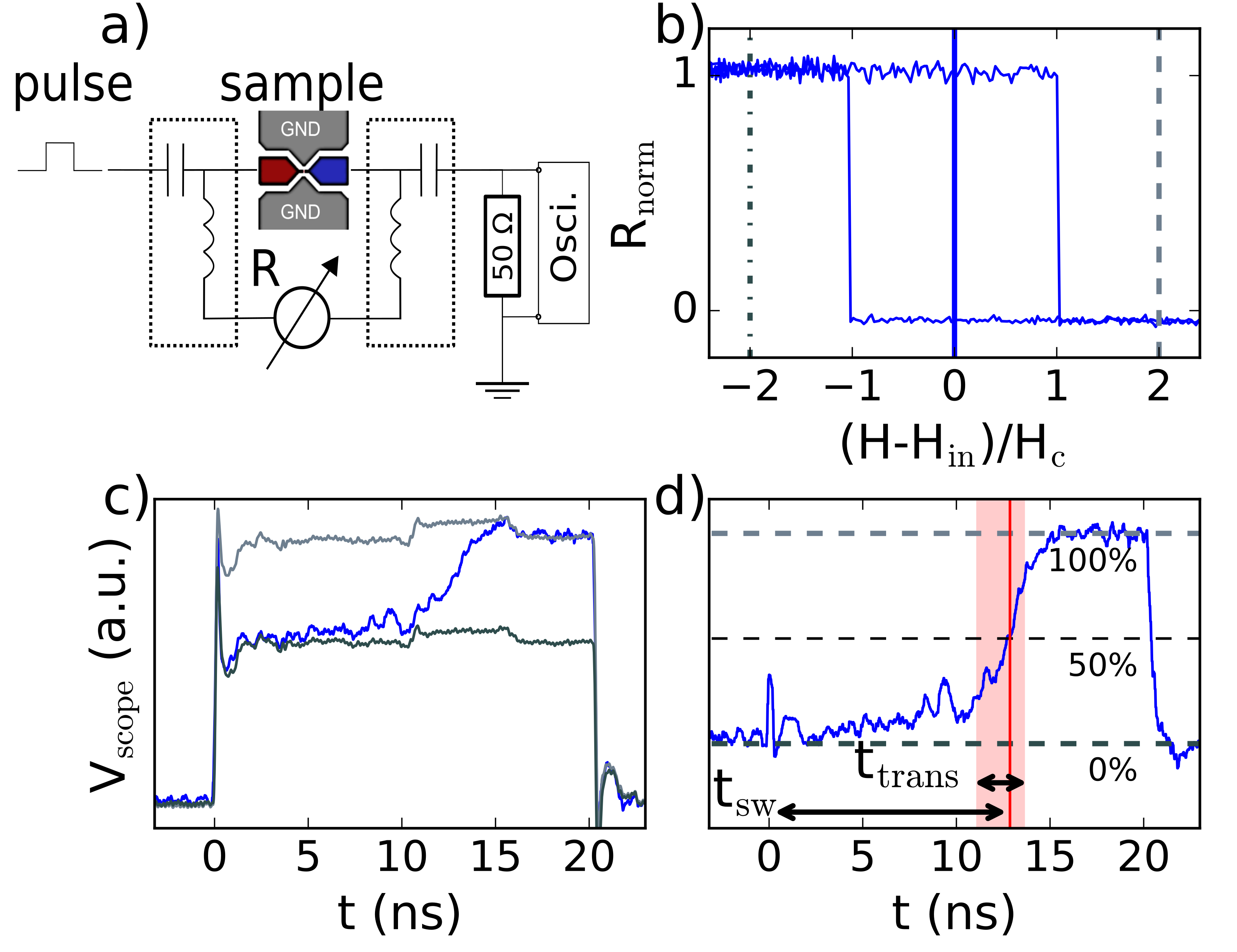}
\caption{(Color online) a) Schematic of the sample and measurement circuit. b) Resistance-field hysteresis loop with dash-dot, dashed and solid lines indicating the fields where the AP and P reference measurements and switching experiment are conducted, respectively. c) Single-shot time-traces of AP reference measurement (dark gray), P reference measurement (light gray) and switching experiment (blue). d) Switched time-trace after subtraction of reference with the times t$_{\mathrm{sw}}$ when the switch occurs (red line) and t$_{\mathrm{trans}}$ the switching duration (semi-transparent red zone) indicated.}
\label{fig:Fig1}
\end{figure}
Figure~\ref{fig:Fig1}\,a) shows a schematic of the measurement circuitry.
The sample is contacted using two rf-probes in a transmission line geometry to enable high-speed measurements. Bias-tees are used to apply dc-currents to measure the junction resistance and to reset the junction's magnetic state after a pulse with a ST torque.  The measurement routine consists of an initialization step during which a small current is applied for a few $\mu$s to set the free layer's magnetization state, a measurement of the pre-pulse dc-resistance, a high-speed pulse measurement and a post-pulse dc-resistance measurement. During the application of the 20\,ns pulse we measure the signal transmitted through the MTJ on a 20\,GHz real-time oscilloscope. We consider our signal to be a measure of the projection of the perpendicular component of the free layer magnetization on the reference layer magnetization. All the measurements were conducted at room temperature.

We use an external field to set the free layer to the parallel (P) and anti-parallel (AP) state, with respect to the reference layer, as indicated indicated in Fig.\,\ref{fig:Fig1}\,b). As a reference, we record voltage traces in both states for positive and negative pulse polarities. The resistance in Fig.\,\ref{fig:Fig1}\,b) has been normalized as $R_{\mathrm{norm}}=(R-R_{\mathrm{P}})/(R_{\mathrm{AP}}-R_{\mathrm{P}})$ and the R(H) curve is shown as a function of a field normalized by the coercivity $H_{\mathrm{c}}$ and shifted by the small internal offset field $H_{\mathrm{in}}$ (associated with dipolar fields from the SAF reference layer) acting on the free layer. The pulsed switching experiment is performed at the center of the hysteresis loop. Fig.\,\ref{fig:Fig1}\,c) shows time-traces of pulses applied to a junction set into the P (dashed line), AP (dotted-dashed line) and at the center of the free layer's hysteresis loop (solid blue line). In this instance, the free layer magnetization switches about 13\,ns after the onset of the pulse, where the blue curve transitions from the AP to the P reference levels. To analyze this data we subtract the reference level of the initial state from the time-traces, resulting in Fig.\,\ref{fig:Fig1}\,d). 
Here, the dashed lines indicate the initial level (0\% switched) the level between AP and P state (50\% switched) and the final state (100\% switched).
These time-traces are used to extract the time when switching occurs (t$_{\mathrm{sw}}$), which we define as the crossing of the 50\% level, indicated by a vertical red line. 
Furthermore, we extract the time needed for magnetization reversal to take place, which we will refer to as transition time (t$_{\mathrm{trans}}$), defined as the time difference between crossing the 25\% and the 75\% level, indicated by the semitransparent red zone.
\begin{figure}
\includegraphics[width=0.5\textwidth]{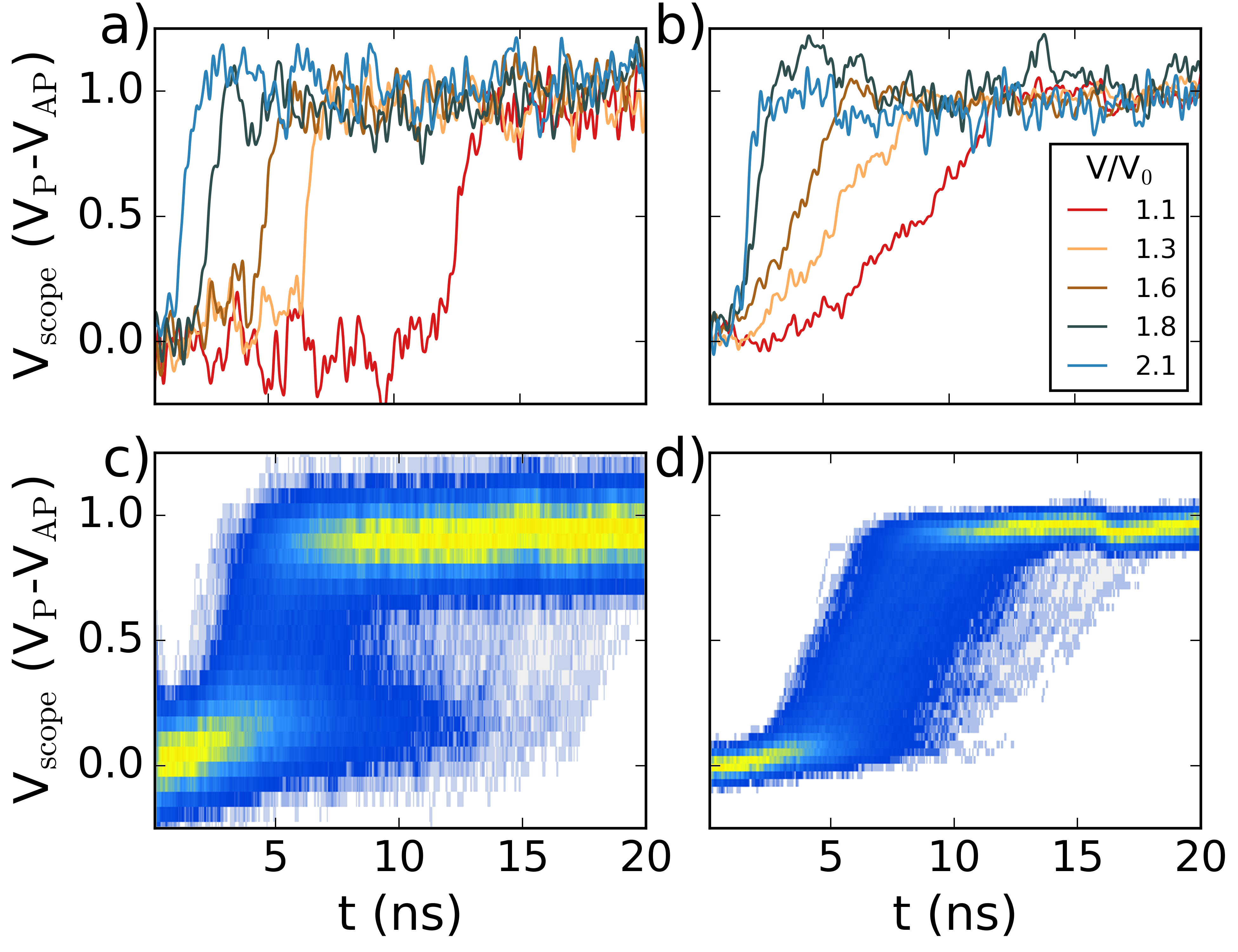} 
\caption{(Color online) Individual switching events of the P$\rightarrow$AP transition for different pulse amplitudes measured on junctions with a) 50\,nm and b) 100\,nm diameter. Overlay of traces for repeated switching events for a fixed pulse amplitude V/V$_0$=1.2 on junctions with c) 50\,nm and d) 100\,nm diameter.}
\label{fig:Fig2}
\end{figure}

We show individual time-traces of magnetization switching from P to AP for different pulse amplitudes on 50 nm (Fig.\,\ref{fig:Fig2}\,a)) and 100\,nm (Fig.\,\ref{fig:Fig2}\,b)) diameter MTJs. The pulse amplitude $V$ has been normalized by $V_0$, the voltage that leads to 50\% switching probability. The time-traces of the 50\,nm junction show a direct reversal with a transition time of about a nanosecond or shorter without a dependence on the voltage pulse amplitude visible to the naked eye. On the other hand, the 100\,nm junction displays a ramp like switching, lasting for several nanoseconds, for the lowest pulse amplitude $V/V_0$=1.1 shown in Fig.\,\ref{fig:Fig2}\,b). The transition time gradually decreases for increasing pulse amplitude until it reaches timescales that are comparable to those observed on the 50\,nm sample. The single time traces in Fig.\,\ref{fig:Fig2}\,a) and b) indicate different timescales for the switching process in the 50 nm and 100 nm diameter junctions.

We record 1000 single-shot measurements for each pulse amplitude. The measurement of the junction's dc-resistance before and after the pulse allows us to categorize the events to have resulted in a switch or a not-switch and sort the measured time-traces accordingly. All time-traces of the switched events for a fixed pulse amplitude V/V$_0$=1.2 are shown in Fig.\,\ref{fig:Fig2}\,c) for the 50\,nm junction and in Fig.\,\ref{fig:Fig2}\,d) for the 100\,nm junction. This pulse amplitude corresponds to greater than 99\% switching probability. Here, the time-traces are overlaid and the resulting image corresponds to what one would see displayed on the fluorescent screen of an oscilloscope after all events, with hot (cool) areas indicating many (few) traces passing through a particular space. The data on the 50 nm diameter sample in Fig.\,\ref{fig:Fig2}\,c) shows a large distribution of the switching times, with switches occurring over almost the entire duration of the pulse, combined with a direct switching mechanism. Despite the wide spread of switching times, the data shows that the majority of switching events occur around 6-7 ns. In contrast to this, Fig.\,\ref{fig:Fig2}\,d) shows that the reversal on the 100 nm junction has a comparatively narrow spread and earlier onset of switching, but the transition time needed for the switching process to occur is larger than on the 50\,nm junction. 
In combination this leads to an average switching time (crossing of the 50\% level) of 6-7\,ns, similar to the 50\,nm junction. This shows a more deterministic onset of the switching on the 100\,nm junction as opposed to the 50\,nm junction where there is a longer incubation time associated with the switching process.
\begin{figure}
\includegraphics[width=0.5\textwidth]{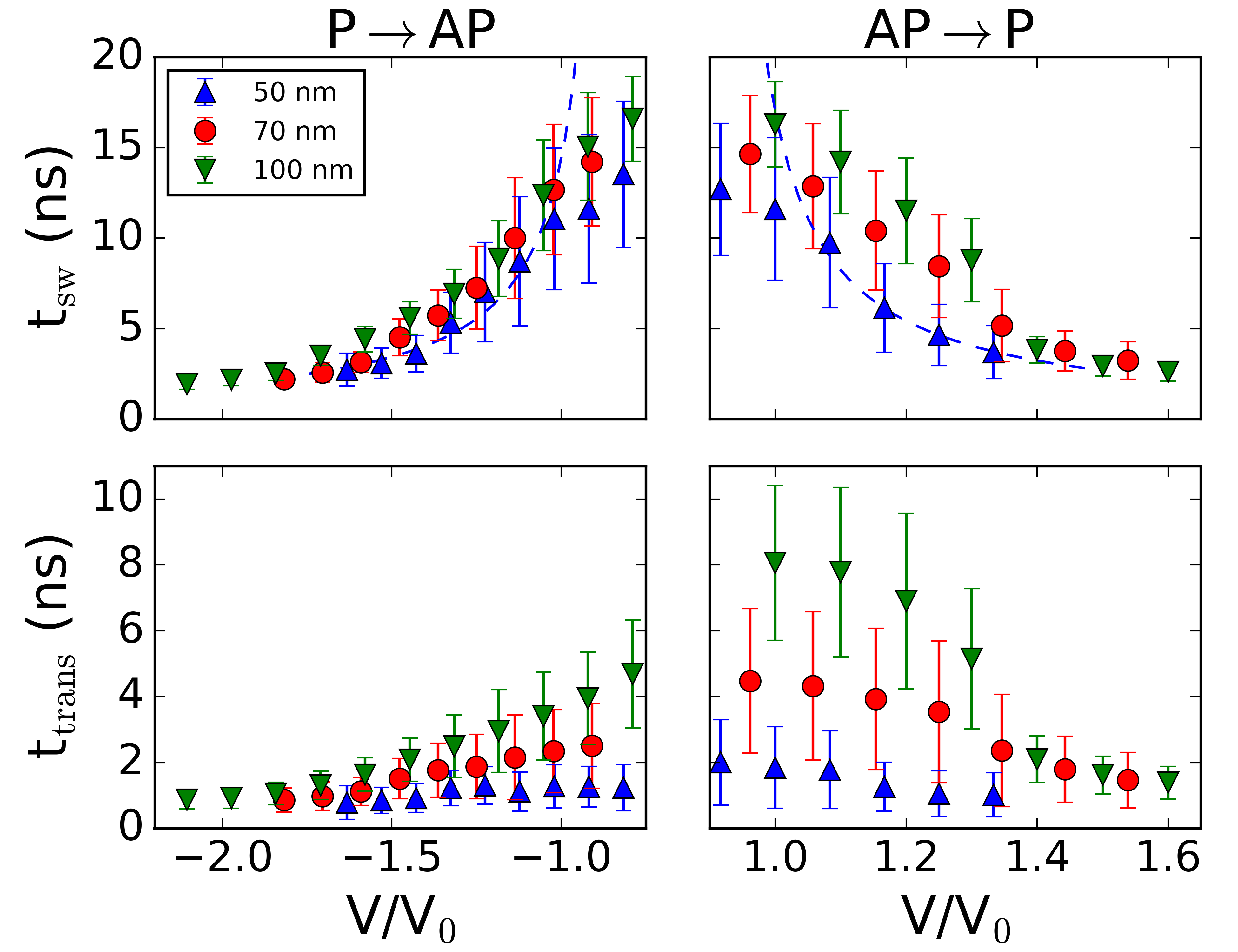}
\caption{(Color online) Mean switching times (t$_{\mathrm{sw}}$) and mean transition times (t$_{\mathrm{trans}}$) as a function of pulse amplitude for three junctions of diameters 50, 70 and 100\,nm. The dashed lines are a fit of 1/t$_{\mathrm{sw}}\propto$V/V$_0$.}
\label{fig:Fig3}
\end{figure}
 
We determine the mean times $t_\mathrm{sw}$ and $t_\mathrm{trans}$ and their standard deviations for all events of a given pulse amplitude and plot them as a function of pulse amplitude in Fig.\,\ref{fig:Fig3}, where green triangles (down), red dots and blue triangles (up) represent the data taken on 100 nm, 70 nm and 50 nm junctions respectively.  The switching times decrease with increasing pulse amplitude and are comparable for junctions of all sizes for the P to AP transition ($V<0$). In contrast to the mean t$_{\mathrm{sw}}$, the standard deviations of the switching times (shown as error bars) differ depending on junction size. The standard deviation of the switching times is larger for smaller junctions. This is consistent with the observation that the switching onset is more spread out in Fig.\,\ref{fig:Fig2}\,a) and c) than in Fig.\,\ref{fig:Fig2}\,b) and d). Overall, the standard deviation decreases with increasing pulse amplitude for all junction sizes. We find an inverse proportionality of the switching time with the applied pulse amplitude. The dashed blue lines are fits of $1/t_{\mathrm{sw}}$ vs. pulse amplitude to the data points for the 50\,nm junctions. Deviations from this form occur as $|V/V_0|$ approaches and becomes less than 1, where the switching probability begins to decrease. The larger size dependence of t$_{\mathrm{sw}}$ for the AP to P ($V>0$) switching direction will be discussed later. 

In contrast to t$_{\mathrm{sw}}$, the transition time shows a clear dependence on junction size. 
For the AP to P transition we find a spread of the mean transition time from 8 ns for the 100 nm junction to 2 ns for the 50 nm junction at a pulse amplitude of $V_0$  (i.e. $V/V_0=1$). The transition time of the 50\,nm junction depends only weakly on pulse amplitude and decreases from 2 to 1 ns with increasing pulse amplitude, while the the transition times of the 70\,nm and 100\,nm junctions decrease strongly to 1.5 ns starting from 4.5 ns and 8 ns respectively. For the P to AP transition, the trend is similar but the initial spread of transition times is smaller, ranging from around 1 ns for the 50 nm junction to over 4 ns for the 100 nm junction at a pulse amplitude of $V_0$. The transition time of the 50\,nm junction shows almost no dependence on pulse amplitude in this switching direction.
The voltage dependence of t$_{\mathrm{trans}}$ in the 100 nm junction and lack thereof in the 50 nm junction, displayed in Fig.\,\ref{fig:Fig3}, reflects the behavior observed already seen in the example time-traces shown in Fig.\,\ref{fig:Fig2}\,a) and b).

\begin{figure}
\includegraphics[width=0.5\textwidth]{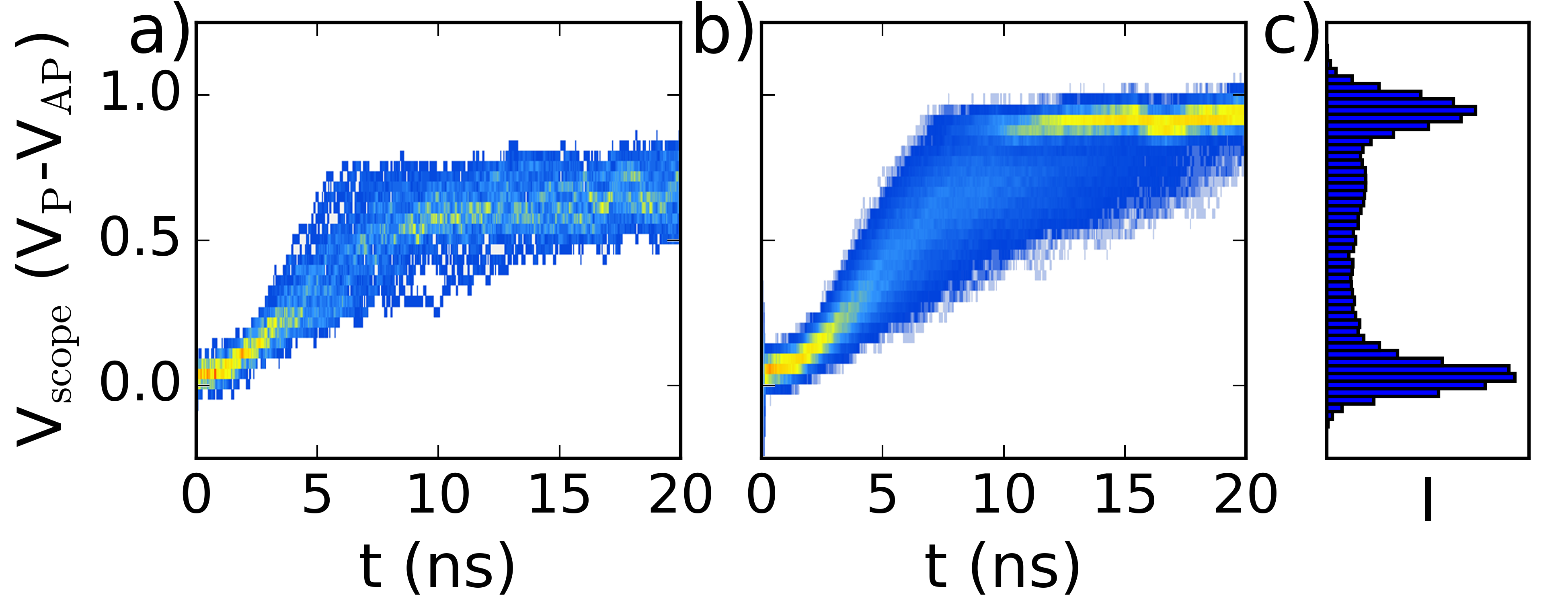}
\caption{(Color online) Overlay of time-traces of the AP$\rightarrow$P transition at V/V$_0$=1.3 sorted by a) not switched and b) switched events. c) Histogram analysis of voltage levels passed through by all the time-traces shown in b); I denotes the intensity at a specific voltage level.}
\label{fig:Fig4}
\end{figure}

We now turn to the transition and switching times of the 100 nm junction for the AP to P transition (i.e. for positive pulse voltages), which are significantly higher than those of the smaller junctions up to 1.3\, $V/V_0$. We find these delayed t$_{\mathrm{sw}}$ and t$_{\mathrm{trans}}$ to be associated with the presence of an intermediate level in the junction resistance. In  Fig.\,\ref{fig:Fig4} we show the density plots of 1000 time-traces at 1.3\,$V/V_0$ pulse amplitude, extracted similarly to those in Fig.\,\ref{fig:Fig2}\,c) and d).  
The time resolved measurements reveal that an intermediate voltage level exists below the level associated with complete magnetization reversal, in which the signal can remain, as can be seen clearly in the not switched events shown in Fig.\,\ref{fig:Fig4}\,a). Here, the applied pulse does not result in a complete switch due to the delay induced by the intermediate state. The time-traces of the switched events, shown in Fig.\,\ref{fig:Fig4}\,b), pass partly through this intermediate level, appearing as a sub-band below the final level, before switching entirely. This reversal path through an intermediate level occurs less with increasing pulse amplitude, leading to an eventual drop in t$_{\mathrm{sw}}$ and t$_{\mathrm{trans}}$. In Fig.\,\ref{fig:Fig4}\,c), a histogram analysis of the time spent at discrete measured voltage amplitude ranges, for the time-traces resulting in a switched magnetization, shows three distinct voltage levels in which the junction remains for longer periods of time: the initial state, the switched state and an intermediate level at about 70\% reversal. 
We find this behavior to be common for the AP to P transition on many of the junctions studied that were 70\,nm or larger in diameter. We note that on no occasion were such intermediate resistance states stable, meaning that after the pulse is turned off the system falls back to either the P or the AP state (i.e. it falls back into these states at least on $\mu$s time scales). Nevertheless, the presence of intermediate states in larger junctions demonstrate a non-uniform magnetization reversal.

The predominance of intermediate resistance states in the AP to P over the opposite transition could be associated with the non-uniformity of the SAF's fringe dipolar field acting on the free layer. While the average dipolar field is compensated by an external field, the fringe field's non uniformity means that in the center of the free layer an (undercompensated) fringe field will favor the parallel orientation, whereas at the edges an (overcompensated) fringe field favors the anti-parallel orientation. When the switching occurs by nucleation of a reversed domain at the element's edge the non-uniformity in the fringe field may hinder (AP starting state) or aid switching (P starting state). The non-uniform fringe field certainly introduces a source of asymmetry between the P to AP and AP to P transitions, as, for example, was reported in perpendicularly magnetized spin valves  \cite{Gopman2012}. 

We will now compare the time needed for magnetization reversal ($t_\mathrm{trans}$) to the timescales expected from an analytical macrospin picture of a uniaxial magnet \cite{Sun2000,Liu2014}. For simplicity we focus primarily on the P to AP transition data. Following Ref.~\cite{Liu2014}, we compute the initial magnetization angle at 300 K before switching, as a function of the size dependent energy barrier. 
Using these initial angles and an overdrive of 1.5 times the critical current, $t_\mathrm{trans}$ is expected to be about 0.7 of the characteristic time scale for the dynamics, $\tau_D = (1+\alpha^2) /(\alpha \gamma \mu_0 H_{k\mathrm{eff}})$. Here, $\alpha$, $\gamma$, H$_{k\mathrm{eff}}$ and $\mu_0$ are the damping constant, the gyromagnetic ratio, the effective perpendicular anisotropy and the vacuum permeability, respectively. We use a vibrating sample magnetometer and thin film ferromagnetic resonance to extract the material parameters damping, magnetization M$_s$=1173\,kA/m and perpendicular anisotropy energy density K$_p$=1026\,kJ/m$^3$. This allows us to compute the size-dependent effective anisotropy $K_\mathrm{{eff}}=K_p-\mu_0 M_s^2 (3N_{zz}-1)$, consisting of the intrinsic perpendicular anisotropy and a size dependent demagnetizing term. We follow Ref.~\cite{Osborn1945} to obtain the size dependent demagnetizing factor $N_{zz}$ analytically and compute it numerically as described in Ref.~\cite{Beleggia2006}, finding that both  methods agree well for the range of sizes of interest here. We find that the theoretical transition time $t_\mathrm{trans}=0.7\cdot \tau_D$ varies from 
6.9\,ns to 7.8\,ns for junctions diameters varying from 50 nm to 100 nm, meaning a weak dependence on the junction size is expected. Note that the estimated transition time varies with overdrive but its dependence on device size does not. This contrasts with the considerable dependence of t$_{\mathrm{trans}}$ on junction size presented in Fig.\,\ref{fig:Fig3} for pulse amplitudes smaller than 1.5 $V/V_0$. The switching times (t$_{\mathrm{sw}}$) predicted from a macrospin model are also larger than those observed.

Taking a simple alternative model of non-uniform magnetization reversal by reversed domain nucleation and expansion by domain wall (DW) motion we can estimate the DW speed assuming that a DW traverses the junction diameter $d$:  $v = d/(2 \cdot \mathrm{t}_{\mathrm{trans}})$. We find similar DW speeds of around 20~m/s for the 70 nm and 100 nm junctions for pulse amplitudes less than 1.5 $V/V_0$. This is consistent with DW velocities reported in Ref.\,\cite{Devolder2016_a}. For the 50\,nm diameter junctions the speed obtained in this amplitude range are higher. 
Therefore, at the low end of our applied pulse amplitude range our data is consistent with a DW mediated reversal, at least on the 70\,nm and 100\,nm diameter junctions.  The size dependent difference in transition time would be explained in this model by domain walls propagating at similar speed across junctions of different size. This is also consistent with the observed ramp-like reversal process, such as that shown in Fig.\,\ref{fig:Fig2}\,b). In comparison, field induced DW propagation speeds in CoFeB thin films of the order of 10~m/s have been observed \cite{Burrowes2013}. 

With increasing voltage pulse amplitude, the mean transition times for junctions of larger size decrease. The highest hypothetical DW speed was measured on the 100\,nm junction and is above 60\,m/s. On the larger end of the voltage range studied we find that transition times show only little variation with junction size. This shows at least qualitative agreement with the macrospin picture for high pulse amplitudes. This may indicate a transition of the reversal mechanism in the larger devices changing from DW mediated reversal at small pulse amplitudes to more coherent macrospin-like reversal at larger amplitudes. 

We now consider the reason why the reversal mechanism changes in junctions that are 50 nm in diameter. This is likely associated with characteristics length scales of the magnetization dynamics. For example, Chaves {\it et al.}~\citep{Chaves2015} computed the energy barrier for thermally activated magnetization reversal of thin disks with perpendicular magnetic anisotropy as a function of their diameter and found two regimes. For diameters smaller than a critical diameter $d_\mathrm{c}$ the reversal occurs by nearly uniform magnetization rotation. Whereas for larger diameters, thermally activated reversal was found to occur by domain wall motion. The critical diameter depends on the free layer's exchange constant $A$ and the size-dependent effective anisotropy $K_{\mathrm{eff}}$: $d_c =(16/\pi) \sqrt{A/K_{\mathrm{eff}}(d)}$. For our free layer materials we estimate that $d_c \simeq 52~\mathrm{nm}$. Thus we are at the critical diameter for uniform thermally activated reversal in our 50~nm junctions. Thus far a model of the characteristic length scale for uniform rotation in the dynamic ST regime is not available. So we can only state at this point that there is a change in magnetization switching dynamics near the length scale at which thermally activated reversal is expected to occur by coherent rotation.

In summary we have studied the switching mechanism in pMTJs as a function of their lateral size and pulse amplitude. We find the mean switching times to be inversely proportional to the pulse amplitude. At low pulse amplitudes we find a considerable dependence of the time needed for the reversal on the junction size, consistent with a reversed domain nucleation and propagation reversal mechanism. However, for larger pulse amplitudes the reversal becomes more abrupt with the transition time decreasing to nanosecond time scales for all junction sizes. Furthermore, we found a dynamical intermediate state in the AP to P transition, which is the origin of a delay in the switching and transition times. Our results thus demonstrate distinct switching mechanisms for pMTJ in the dynamic ST voltage pulse limit, and highlight the need for more sophisticated models of their switching characteristics in this limit.

This research is supported by Spin Transfer Technologies Inc.

\bibliography{pMTJAK} 
\end{document}